# Screening effect of a magnetically soft shell of single-domain ferromagnetic nanoparticles

Vladimir P. Savin[a*], Yury A. Koksharov[a,b]

[a]Faculty of Physics, M.V.Lomonosov Moscow State University, Leninskie Gory, Moscow, 119991, Russia

[b]Kotelnikov Institute of Radio-Engineering and Electronics, Russian Academy of Sciences, Moscow, 125009 Russia

*Corresponding author.

e-mail: svladimir9915@gmail.com (Savin V.P.)

**Abstract**

The possible manifestations of magnetic screening effects were theoretically investigated for a particle with a ferromagnetic single-domain core and a magnetically soft shell. The exact solution of the Laplace equation gave analytical formulas for the magnetic field created by the particle placed in a uniform external magnetic field. The system of non-interacting randomly oriented particles was investigated in the Langevin model. The effects of the shell on individual and thermodynamically averaged magnetic properties of the particles were described in a single way using the screening factor $A$. The screening effects are appeared in a decrease in: (i) the demagnetization factor of the particle core; (ii) the total dipole moment and, accordingly, the external magnetic field created by the particle; (iii) the effective magnetic moment in isothermal magnetization curves and in the Curie law. The magnetically soft shell of the particle also changes the behavior of the particle system in high magnetic fields: the magnetization does not reach saturation, and the differential magnetic susceptibility tends to a non-zero value.

# 1. Introduction

Screening effects are not uncommon in physics. For example, electric charge is screened in conductors due to a redistribution of free electrons in space. The magnetic field or magnetic moment can be screened by superconducting currents in the Meissner effect. Magnetic moments of ferromagnetic particles embedded in a superconductor can be screened by the spins of Cooper pairs [1]. Magnetic screens [2] are used for shielding of electronic devices, to investigate magnetic fields generated by biological samples [3].

Are there any possible screening effects in nano-systems? Many types of nanoobjects are considered in the literature. They differ in size, shape, morphology and the nature of interparticle interactions [4-24]. The physical, chemical and biological properties of nanoparticles are also very diverse [4-11], [22]. By changing the internal structure of the nanoparticles, one could try to achieve their desired characteristics. A popular model of the internal structure of nanoparticles is the "core-shell" model [12-20], [25, 26]. In a nanoparticle the properties of the surface and internal atoms differ, even if the chemical nature and crystal structure of the core and the surface layer are the same. Experiments have shown that artificial modification in the surface composition leads to a significant change in the properties of nanoparticles [19, 20, 27, 28].

Magnetic nanoparticles are particularly interesting from a fundamental and applied point of view, since they can be controlled using an external magnetic field [29, 30]. There are many experimental [13, 14, 17, 18, 19], [31-36] and theoretical [21, 35], [37-40] studies of magnetic particles with a core-shell structure. The magnetic disordering in the nanoparticle shell have been used to explain the phenomenon of decreasing magnetization of nanoparticles compared to bulk materials [41, 42]. Other mechanisms of this phenomenon are also discussed, for example, the presence of antiphase boundaries in the core of the nanoparticle [43, 44].

Theoretical analysis of the magnetic properties of shell-less single-domain nanoparticles are usually based on the classical Langevin [45], Stoner-Wolfart [46], Néel [47] and more advanced models (see e.g. [48]). The analysis of the magnetic properties of core-shell nanoparticles is a much more complex task and has so far been carried out only by numerical methods [34, 37, 39, 40]. In this paper, we have considered a fairly simple model of a magnetic core-shell nanoparticle, which provides an accurate analytical solution. The abbreviation CSHS (core – shell hard – soft) is used, since the single-domain core is hard ferromagnet and the shell is soft ferromagnet in this model.

It is of interest to study the system of the CSHS nanoparticle in classical models (Langevin, Stoner-Wohlfarth, etc.) and compare it with magnetic nanoparticles without a shell. A hollow sphere with a magnetically soft spherical shell and a magnetic point dipole inside was investigated in the Runcorn model [49, 50]. A drawback of the Runcorn's original paper (ignoring the demagnetization field of the shell) was fixed in the work [51]. In the Runcorn model, the shell acts as a self-screening element. It is not possible to use directly results of [49-51], since there is no external magnetic field in the Runcorn model. In addition, there is a difference between the internal magnetic fields in our model and in the Runcorn model. Thus, we obtained a general analytical solution to the problem of magnetic properties of the CSHS particle in the external magnetic field. The solution can be interpreted as "screening": (i) the shell screens the outer space from the core magnetic field; and (ii) the shell screens the core from the external magnetic field. Finally, we calculated the thermodynamically equilibrium magnetic properties of the CSHS particle system in the Langevin model and found manifestations of magnetic screening in the superparamagnetic behavior of the CSHS particle ensemble.

## 2. The CSHS particle model

Let us consider a non-magnetic medium with a spherical particle embedded in it and placed in an external homogeneous field $\vec{H}_0$ (Fig.1). The particle consists of a single-domain hard ferromagnetic core of radius $a$ and a soft magnetic shell of thickness $d = b - a$, where $b$ is the particle radius. $\vec{M}_0$ is the uniform magnetization of the core and $\mu$ is the permeability of the homogeneous shell.

The total magnetic moment $\vec{m}$ of the CSHS particle can be represented as a sum of magnetic moments $\vec{m}_0$ of the core and $\vec{m}_{sh}$ of the shell:

$$\vec{m} = \vec{m}_0 + \vec{m}_{sh} \qquad (1)$$

Generally, the magnetic moments of the core and the shell are non-parallel. The shell magnetization $\vec{M}_{sh}$ is non-homogeneous since it depends on the dipole-like magnetic field of the core in the region II (Fig.1).

## 3. Results

### 3.1. The solution of the Laplace equation

In order to find the magnetic moment of the CSHS particle in the external homogeneous field $\vec{H}_0$ let us consider, without loss of generality, that the magnetic moment of the core is directed along the $z$ axis (Fig.1):

$$\vec{m}_0 = m_0 \vec{e}_z = V_0 M_0 \vec{e}_z, \tag{2}$$

where $V_0 = 4\pi a^3/3$ is the core volume.

The direction of the external magnetic field $\vec{H}_0$ is determined by the polar $\theta_H$ and azimuth $\varphi_H$ angles of a spherical coordinate system:

$$\frac{\vec{H}_0}{|\vec{H}_0|} = (sin\theta_H cos\varphi_H, sin\theta_H sin\varphi_H, cos\theta_H). \tag{3}$$

In order to find the magnetic moment of the shell, it is necessary to determine the field in region II (see Fig.1) by solving the boundary value problem. In a region of constant permeability $\mu$ and free of electrical currents, the magnetic field strength is:

$$\vec{H} = -grad\psi, \tag{4}$$

where $\psi$ is the magnetic potential, which satisfies Laplace's equation:

$$\Delta\psi = 0. \tag{5}$$

Taking into account the boundary conditions at two spherical surfaces $r = a$ and $r = b$ leads to four equations:

$$\psi^I|_{r=a} = \psi^{II}|_{r=a}; \tag{6}$$

$$\psi^{II}|_{r=b} = \psi^{III}|_{r=b}; \tag{7}$$

$$\left(-\frac{\partial\psi^I}{\partial r} + 4\pi M_{0n}\right)\bigg|_{r=a} = -\mu\frac{\partial\psi^{II}}{\partial r}\bigg|_{r=a}; \tag{8}$$

$$\mu\frac{\partial\psi^{II}}{\partial r}\bigg|_{r=b} = \frac{\partial\psi^{III}}{\partial r}\bigg|_{r=b}, \tag{9}$$

where $M_{0n}$ is a normal component of the magnetization vector $\vec{M}_0$ to the spherical surface with $r = a$; $r = \sqrt{x^2 + y^2 + z^2}$.

Using Eq. (2), we can express the normal component of the core magnetization as

$$M_{0n} = M_0 cos\theta. \tag{10}$$

The potential $\psi$ can be expressed in terms of spherical harmonics. Taking into consideration that $\psi$ is finite near the origin ($r = 0$) and vanishes at $r \to \infty$, we can write:

Region I ($0 < r < a$):

$$\psi^I(\vec{r}) = \psi^I_{core}(\vec{r}) + \sum_{n=0}^{\infty}\sum_{m=0}^{n} r^n P_n^{(m)}(cos\theta)\{A^I_{nm}cosm\varphi + B^I_{nm}sinm\varphi\} \qquad (11)$$

Region II ($a < r < b$):

$$\psi^{II}(\vec{r}) = \psi^{II}_{core}(\vec{r}) + \sum_{n=0}^{\infty}\sum_{m=0}^{n} \left(\frac{r^{2n+1} - a^{2n+1}}{r^{n+1}}\right) P_n^{(m)}(cos\theta)\{A^{II}_{nm}cosm\varphi + B^{II}_{nm}sinm\varphi\} +$$

$$+ \sum_{n=0}^{\infty}\sum_{m=0}^{n} \left(\frac{b^{2n+1} - r^{2n+1}}{r^{n+1}}\right) P_n^{(m)}(cos\theta)\{C^{II}_{nm}cosm\varphi + D^{II}_{nm}sinm\varphi\}; \qquad (12)$$

Region III ($b < r$):

$$\psi^{III}(\vec{r}) = \psi^{III}_H(\vec{r}) + \sum_{n=0}^{\infty}\sum_{m=0}^{n} \frac{1}{r^{n+1}} P_n^{(m)}(cos\theta)\{A^{III}_{nm}cosm\varphi + B^{III}_{nm}sinm\varphi\}, \qquad (13)$$

where $\psi^I_{core}(\vec{r})$ and $\psi^{II}_{core}(\vec{r})$ are the magnetic potential produced by the core in regions I and II, respectively; $\psi^{III}_H(\vec{r})$ is the magnetic potential of the external field $\vec{H}_0$; $P_n^{(m)}(cos\theta)$ are the associated Legendre functions of the first kind. $A^I_{nm}$, $B^I_{nm}$, $A^{II}_{nm}$, $B^{II}_{nm}$, $C^{II}_{nm}$, $D^{II}_{nm}$, $A^{III}_{nm}$, $B^{III}_{nm}$ are constant coefficients to be found.

In region I, the magnetic potential of the core is equal to:

$$\psi^I_{core}(\vec{r}) = \frac{4\pi M_0}{3} r cos\theta = \frac{m_0}{a^3} r cos\theta. \qquad (14)$$

In region II, the magnetic potential of the core corresponds to the field of a point dipole:

$$\psi^{II}_{core}(\vec{r}) = \frac{4\pi M_0}{3}\frac{a^3}{r^2} cos\theta = \frac{m_0}{r^2} cos\theta. \qquad (15)$$

The potential of the external homogeneous field can be written as:

$$\psi^{III}_H(\vec{r}) = -r(H_{0x}sin\theta cos\varphi + H_{0y}sin\theta sin\varphi + H_{0z}cos\theta). \qquad (16)$$

where $H_{0x}, H_{0y}, H_{0z}$ are the $x, y, z$ vector components of the vector $\vec{H}_0$.

Note that according to the boundary conditions (6) - (9), the terms with indices $n > 1$ vanish in the equations (11)-(13). Taking into account $P_1^{(0)}(cos\theta) = cos\theta$, $P_1^{(1)}(cos\theta) = sin\theta$ and (14) – (16) we deduce from (11)-(13):

$$\psi_I(\vec{r}) = \frac{m_0}{a^3} r cos\theta + r(A^I_{10}cos\theta + A^I_{11}sin\theta cos\varphi + B^I_{11}sin\theta sin\varphi); \qquad (17)$$

$$\psi_{II}(\vec{r}) = \frac{m_0}{r^2}\cos\theta + \left(\frac{r^3 - a^3}{r^2}\right)(A_{10}^{II}\cos\theta + A_{11}^{II}\sin\theta\cos\varphi + B_{11}^{II}\sin\theta\sin\varphi) +$$

$$+ \left(\frac{b^3 - r^3}{r^2}\right)(C_{10}^{II}\cos\theta + C_{11}^{II}\sin\theta\cos\varphi + D_{11}^{II}\sin\theta\sin\varphi); \tag{18}$$

$$\psi_{III}(\vec{r}) = -r(H_{0x}\sin\theta\cos\varphi + H_{0y}\sin\theta\sin\varphi + H_{0z}\cos\theta) +$$

$$+ \frac{1}{r^2}(A_{10}^{III}\cos\theta + A_{11}^{III}\sin\theta\cos\varphi + B_{11}^{III}\sin\theta\sin\varphi). \tag{19}$$

For the convenience of further presentation let us introduce the "screening factor" $A$:

$$A = \frac{1}{1 + \frac{2}{9}\frac{(\mu - 1)^2}{\mu}\frac{(b^3 - a^3)}{b^3}} = \frac{1}{1 + \frac{2}{9}\frac{(\mu - 1)^2}{\mu}\frac{V_{sh}}{V_p}}. \tag{20}$$

where $V_{sh} = 4\pi(b^3 - a^3)/3$ is the shell volume and $V_p = 4\pi b^3/3$ is the particle total volume.

The value of $A$ depends on the magnetic (via $\mu$) and geometrical properties (via $a$ and $b$) of a particle and ranges from 0 (for $\mu \gg 1$) to 1 (for $b = a$ or $\mu = 1$).

When the boundary conditions (6) – (9) are applied to (17) – (19), we obtain three independent sets of linear equations for the coefficients $A_{10}^{I}$, $A_{11}^{I}$, $B_{11}^{I}$, $A_{10}^{II}$, $A_{11}^{II}$, $B_{11}^{II}$, $C_{10}^{II}$, $C_{11}^{II}$, $D_{11}^{II}$, $A_{10}^{III}$, $A_{11}^{III}$, $B_{11}^{III}$. The sets of linear equations (A.1) - (A.3) and explicit expressions (A.4) - (A.15) for coefficients $A_{10}^{I}$, $A_{11}^{I}$, $B_{11}^{I}$, $A_{10}^{II}$, $A_{11}^{II}$, $B_{11}^{II}$, $C_{10}^{II}$, $C_{11}^{II}$, $D_{11}^{II}$, $A_{10}^{III}$, $A_{11}^{III}$, $B_{11}^{III}$ in regions I, II and III are given in Appendix A.

The formulas (17)-(19) and (A.4) - (A.15) give an exact solution to the Laplace equation for the scalar magnetic potential in the problem under consideration.

### 3.2. Magnetic properties of the CSHS particle

Taking into account Eqs. (4), (17)-(19) and using (A.4) - (A.15), we obtained general formulas for the magnetic field strength in the core ($\vec{H}_{I}$), in the shell ($\vec{H}_{II}$) and out of the particle ($\vec{H}_{III}$):

$$\vec{H}_I = A\vec{H}_0 - \frac{\vec{m}_0}{a^3}\left(1 - (1 - A)\frac{\mu + 2}{\mu - 1}\right); \tag{21}$$

$$\vec{H}_{II} = \vec{H}_{core}\left[1 + A\frac{2(\mu - 1)^2}{9\mu}\left(\frac{a^3}{b^3} - \frac{\mu + 2}{\mu - 1}\right)\right] +$$

$$+A\frac{2\mu+1}{3\mu}\vec{H}_0 - A\frac{(\mu-1)}{3\mu}a^3\left(\frac{3\vec{r}(\vec{H}_0\vec{r})-\vec{H}_0 r^2}{r^5}\right) - A\frac{2(\mu-1)}{3\mu}\frac{\vec{m}_0}{b^3}; \qquad (22)$$

$$\vec{H}_{III} = \vec{H}_0 + \frac{3\vec{r}(\vec{m}\vec{r})-\vec{m}r^2}{r^5}; \qquad (23)$$

where $\vec{H}_{core}$ in (22) is the field of a point dipole which has the magnetic moment $\vec{m}_0$:

$$\vec{H}_{core} = \frac{3\vec{r}(\vec{m}_0\vec{r})-\vec{m}_0 r^2}{r^5}. \qquad (24)$$

The same dipole field is created by a uniformly magnetized sphere in the external space (see Eq.(5.106) in [52]). The vector $\vec{m}$ in (23) is the notation for the sum:

$$\vec{m} = A\vec{m}_0 + (1-A)\frac{2\mu+1}{2\mu-2}b^3\vec{H}_0. \qquad (25)$$

It is equal to the total magnetic moment $\vec{m}$ of the particle (see Eq.(1)). Indeed, the total magnetic moment of the shell can be obtained by integrating the shell magnetization:

$$\vec{M}_{sh} = \frac{(\mu-1)}{4\pi}\vec{H}_{II} \qquad (26)$$

over the volume of the shell:

$$\vec{m}_{sh} = \int_{V_{sh}} \vec{M}_{sh} dV \qquad (27)$$

Substituting Eqs. (22), (26) in (27) and integrating over volume $V_{sh}$ we obtained

$$\vec{m}_{sh} = (1-A)\left(\frac{2\mu+1}{2\mu-2}b^3\vec{H}_0 - \vec{m}_0\right) \qquad (28)$$

The sum of $\vec{m}_{sh}$ and $\vec{m}_0$ gives $\vec{m}$ in (25). The first term $A\vec{m}_0$ in (25) can be interpreted as the screened magnetic moment of the core. Then the second term in (25) represents induced magnetic moment of the shell, which is proportional to the external field and the particle volume. This term increases with decreasing $A$.

The field (21) in the core can be considered as a superposition of the external field, reduced by a factor of $A$, and the demagnetization field. The demagnetization factor is reduced by a term depending on the parameter $A$.

The bulky sum (22) contains several contributions: (i) the dipole field $\vec{H}_{core}$ modified by a multiplier (see straight brackets in (22)); (ii) the external field reduced by $A(2\mu+1)/(3\mu)$ factor;

(iii) the dipole-like field depending on the external field; (iv) the demagnetization-like homogeneous field depending on the core magnetic moment $\vec{m}_0$ and reduced by $2A(\mu-1)/(3\mu b^3)$ factor. Note that dipole-like terms in (22) do not give contribution to the shell magnetic moment calculated by (27).

Fig.2 shows inhomogeneous distribution of the magnetization $\vec{M}_{sh}$ (Eq.(26)) inside the shell (near internal and external surfaces), in the case of $\vec{m}_0 = const$. If $\vec{H}_0 = 0$, then the distribution is axisymmetric with the onion-like magnetic structure [53]. If $\vec{H}_0 < \vec{M}_0$, then the magnetic structure is not affected significantly by the external magnetic field. In high magnetic fields, the shell magnetization tends to be parallel to $\vec{H}_0$.

The field (23) in space outside the particle is the sum of the undisturbed external field and the point dipole field of the magnetic moment $\vec{m}$ given by Eq.(25). Two special cases of Eq.(21) and Eq.(23) are examined in Appendix B.

### 3.3. The CSHS particle system in the Langevin model

### 3.3.1. The energy of the CSHS particle in the external magnetic field

The Langevin model takes into account thermal fluctuations of the particle magnetic moment. In case of the particle without the shell the magnetic moment direction only fluctuates. For the CSHS particle the absolute value of the total magnetic moment (25) can also change.

It is convenient to write the magnetostatic energy of a particle as the sum of two terms:

$$W = W_1 + W_2. \tag{29}$$

The first term in (29) is the interaction energy of the permanent core moment $\vec{m}_0$ with the external homogeneous field $\vec{H}_0$:

$$W_1 = -(\vec{m}_0 \vec{H}_0) = -m_0 H_0 cos\theta_H, \tag{30}$$

The second term is the energy of the shell's interaction with the magnetic field created by all static, unchangeable external sources (see Eq.(32.5) in [54]):

$$W_2 = -\frac{1}{2} \int_{V_S} \left( \vec{M}_{sh}(\vec{H}_0 + \vec{H}_{core}) \right) dV \tag{31}$$

where $\vec{H}_{core}$ is the magnetic field of the core, $\vec{M}_{sh}$ is the magnetization of the shell. The integration in (31) is extended over the volume of the shell.

Integrating (31) and taking (22), (24), (26) into account, we obtain:

$$W_2 = -\frac{1}{2}(1-A)\left[\frac{2\mu+1}{2\mu-2}b^3 H_0^2 - 2(\vec{m}_0\vec{H}_0) + \frac{\mu+2}{\mu-1}\frac{m_0^2}{a^3}\right]; \quad (32)$$

The total energy (29) takes the form:

$$W = -\frac{1}{2}(1-A)\frac{2\mu+1}{2\mu-2}b^3 H_0^2 - A m_0 \vec{H}_0 - \frac{1}{2}(1-A)\frac{\mu+2}{\mu-1}\frac{m_0^2}{a^3}. \quad (33)$$

The first and second terms in (33) can be interpreted as the energy of interaction of the induced and permanent components (see Eq.(25)) of the particle magnetic moment with an external magnetic field $\vec{H}_0$. The last term in (33) is related to the magnetic interaction of the core and shell in absence of the external field.

The special case of (33) can be calculated by putting $a = 0$ and $\vec{m}_0 = 0$:

$$W|_{a=0} = -\frac{b^3 H_0^2 (\mu-1)}{2(\mu+2)}. \quad (34)$$

Eq. (34) coincides with Eq. (5.115) in [52] for the free energy of a homogeneous soft magnetic sphere.

### 3.4.2. The equilibrium magnetic moment of the CSHS particles in the Langevin model

Let us now consider of the thermodynamically equilibrium ensemble of identical noninteracting CSHS particles in the external homogeneous field $\vec{H}_0$. The total magnetic energy of the system is equal to the sum of the energies $W$ of individual magnetic particles.

For the sake of simplicity in notation we can reduce the energy to the form:

$$-\beta W(\theta_H) = A\xi \cos\theta_H + B, \quad (35)$$

where

$$\beta = 1/k_B T; \quad (36)$$

$$\xi = \frac{m_0 H_0}{k_B T}; \quad (37)$$

$$B = \frac{(1-A)}{2k_B T}\left[\frac{2\mu+1}{2\mu-2}b^3 H_0^2 + \frac{\mu+2}{\mu-1}\frac{m_0^2}{a^3}\right]; \quad (38)$$

$T$ is the absolute temperature, $k_B$ is the Boltzmann constant. We have indicated the dependence of the energy $W$ on the angle $\theta_H$ in Eq. (35).

The statistical independence of non-interacting magnetic particles makes it possible to obtain the thermodynamic parameters of the system by summing up all possible states of a single particle. Each state of a particle is determined by the orientation of the magnetic moment $\vec{m}$ with respect to the external field $\vec{H}_0$. The orientation is determined by the spherical angles $\theta_H$ (see Fig.1) and $\varphi_H$ (not shown in Fig.1). In the Langevin model the energy $W$ does not depend on $\varphi_H$.

The partition function related to the particle energy can be defined as (see Eq.(2.9) in [48]):

$$Z = \frac{1}{2\pi} \int_0^\pi d\theta_H \sin\theta_H \int_0^{2\pi} d\varphi_H \exp[-\beta W(\theta_H)] \tag{39}$$

The equilibrium probability distribution of magnetic moment orientations is given by:

$$P(\theta_H) = Z^{-1} \exp[-\beta W(\theta_H)] \tag{40}$$

The thermodynamical average of the projection $\langle m_{\vec{H}_0} \rangle$ of the particle magnetic moment $\vec{m}$ on the external field $\vec{H}_0$ is equal to

$$\langle m_{\vec{H}_0} \rangle = \frac{1}{2\pi} \int_0^{2\pi} \sin\theta_H \, d\theta_H \int_0^\pi d\varphi_H \, m_{\vec{H}_0} P(\theta_H) \tag{41}$$

where $m_{\vec{H}_0}$ can be written as (see Eq.(25)):

$$m_{\vec{H}_0} = m_0 A \cos\theta_H + (1-A) \frac{2\mu+1}{2\mu-2} b^3 H_0, \tag{42}$$

For the CSHS particle without a shell, Eq.(42) reduces to the Langevin expression:

$$m_{\vec{H}_0} = m_0 \cos\theta_H \tag{43}$$

Integrating (39), we obtain:

$$Z = \frac{2e^B}{A\xi} \sinh A\xi \tag{44}$$

Using Eqs.(35), (39)-(44), we can write an expression for $\langle m_{\vec{H}_0} \rangle$:

$$\langle m_{\vec{H}_0} \rangle = m_0 \frac{\partial \ln Z}{\partial \xi} + (1-A) \frac{2\mu+1}{2\mu-2} b^3 H_0 = A m_0 L(A\xi) + (1-A) \frac{2\mu+1}{2\mu-2} b^3 H_0 \tag{45}$$

where $L(z) = cth(z) - 1/z$ is the Langevin function;

To check Eq.(45) we can consider two limiting cases:

a) If there is no shell ($b = a$, $A = 1$), we get the Langevin's result:

$$\langle m_{\vec{H}_0}\rangle\big|_{a=b} = m_0 L(\xi); \tag{46}$$

b) If $a = 0$ we obtain the well-known result for a soft magnetic sphere in the external magnetic field $\vec{H}_0$ (see (5.115) in [52]):

$$\langle m_{\vec{H}_0}\rangle\big|_{a=0} = \frac{\mu - 1}{\mu + 2} b^3 H_0, \tag{47}$$

### 3.4.3. Specific superparamagnetic properties of CSHS particles

Fig.3 shows the normalized dependences $\langle m_{\vec{H}_0}\rangle$ versus $H_0$ for CSHS particles with different values of ratio $a/b$. To illustrate these dependencies, it is convenient to define the normalizing coefficient as the maximum magnetic moment $m_{0,max}$ related to a single-domain particle with radius $b$:

$$m_{0,max} = \frac{4\pi b^3}{3} M_0 \tag{48}$$

If the core volume fraction ($a^3/b^3$) decreases (see Fig.3), the type of the field dependences $\langle m_{\vec{H}_0}\rangle(H_0)$ changes from superparamagnetic (non-linear) to paramagnetic (linear). Such behavior proves a correctness of Eq.(45), since parameters $A$, $\xi \sim m_0 \sim a^3$ reduce and the second term begins to prevail with decreasing $a/b$. Fig.3 also demonstrates the screening effect of the particles shell, which manifests in a reduction in the average magnetic moment $\langle m_{\vec{H}_0}\rangle$.

Taking into account the asymptotical formulas $L(x)_{x\ll 1} \approx x/3$ and $L(x)_{x\gg 1} \approx 1 - 1/x$, we can rewrite the Eq. (45) for the case of $A\xi \ll 1$:

$$\frac{\langle m_{\vec{H}_0}\rangle}{Am_0} = \frac{A\xi}{3} + \frac{(1-A)}{A^2}\frac{2\mu + 1}{2\mu - 2}b^3\frac{k_B T}{m_0^2}A\xi \tag{49}$$

and for the case of $A\xi \gg 1$:

$$\frac{\langle m_{\vec{H}_0}\rangle}{Am_0} = 1 - \frac{1}{A\xi} + \frac{(1-A)}{A^2}\frac{2\mu + 1}{2\mu - 2}b^3\frac{k_B T}{m_0^2}A\xi. \tag{50}$$

Fig.4a shows $\langle m_{\vec{H}_0}\rangle$ (Eq.(45)) normalizing by $Am_0$ versus $A\xi$. The use of the screening parameter $A$ as a scale factor makes it possible to clearly observe the influence of the shell on the isothermal field dependences of the average magnetic moment. In addition, the choose of specific coordinates in Fig.4a allow us to find universality in the behavior of the isothermal field curves.

Firstly, the tangent of the slop angle $\beta$ of the curves for $A\xi \ll 1$ is nearly identical for all values of $A$ ($\mathrm{tg}\,\beta = 1/3$). This is true if the second term in (49) is small in comparison with the first one. This condition requires that $M_0 \gg 100\,G$ at room temperature and not too small values of $A$. Secondly, the tangents drawn at $A\xi \gg 1$ intersect at one point of the ordinate axis (see Fig.4a and Eq.(50), in which we neglect $1/A\xi$). The tangent of the slop angle $\alpha$ is proportional to $(1-A)/A^2$ and it increases with decrease of $A$, in accordance with (50). Thus, in these specific coordinates, the impact of the shell is negligible at low magnetic fields and significant at high magnetic fields. It is possible to apply the universality observed in Fig.4a to find the parameter $A$ from the experimental data. To do this, it is necessary to vary the value of $A$, achieving the tangent behavior shown in Fig.4a.

The dependencies of $\langle m_{\vec{H}_0}\rangle/m_0$ versus $\xi$ are shown in Fig.4b. Under isothermal conditions (close to room temperatures), if $\xi \ll 1$ and $M_0 \gg 100\,G$, the tangent of the angle of inclination $\gamma$ increases with increasing $A$ as $A^2/3$ (see Eq.(49)). Such behavior corresponds to a weakening of the screening effect with the increase in the value of $A$. If $\xi \gg 1$, the value of $\langle m_{\vec{H}_0}\rangle/m_0$ grows slowly as $A - 1/\xi$. This is true if $\xi$ is not very large, so that the third term in Eq.(50) can be neglected. Hence, under these conditions the factor $A$ reduces high-field average magnetic moment of CSHS particles in comparison with the shell-less single domain particles.

### 3.4.4. The Curie law for CSHS particles

At low fields, the core contribution to the total magnetic moment of the particle can be written in Curie-like form (see the first term in Eq.(49)):

$$m_c = A^2 \frac{m_0^2 H_0}{3k_B T} \sim \frac{1}{T} \qquad (51)$$

Fig.5 shows the difference between temperature dependences of the average magnetic moment in Eq. (45) and Eq. (51), which demonstrates the effect of the shell on the fulfillment of the Curie law for CSHS particles. The insert in Fig.5 shows the modulus of the difference $\varepsilon$ (normalized by $m_0$) between the temperature dependences (45) and (51). The Curie law does not hold well at low temperatures, both for particles with and without a shell. For the particles without a shell ($A = 1$, see curve (1) in Fig.5) the average magnetic moment (Eq.(46)) satisfies the Curie law well only at $T > 100\,K$, and if $T \to \infty$, then $\varepsilon \to 0$. For CSHS particles ($A < 1$, see curves (2), (3) in Fig.5), Eq.(45) becomes satisfying the Curie law at lower temperatures compared to Eq.(46). This is due to the presence of the multiplier $A$ in the argument of the Langevin function in Eq.(45) and Eq.(46). It expands the area in which the argument is small. At high temperatures the error $\varepsilon$ for CSHS particles increases. This deviation is determined by the second term in Eq. (45), which depends on $A$ and $H_0$. The value of $\varepsilon$, which is constant at high temperatures (insert in Fig.5), can be used to find the screening parameter $A$ based on experiments.

### 3.4.5. The magnetic susceptibility of CSHS particles

Taking into account Eq. (45) and the relation $(L'(x) = 1 - \frac{2}{x}L(x) - L^2(x))$ it is possible to obtain an expression for the differential magnetic susceptibility per particle:

$$\frac{\chi_T}{n} = \frac{\partial \langle m_{\vec{H}_0} \rangle}{\partial H_0} = \frac{m_0^2 A^2}{k_B T}\left(1 - \frac{2}{A\xi}L(A\xi) - L^2(A\xi)\right) + (1-A)\frac{2\mu+1}{2\mu-2}b^3, \qquad (52)$$

where $n$ is the volume concentration of CSHS particles in non-magnetic medium.

Taking into account the asymptotical formulas $L(x)_{x \ll 1} \approx x/3 - x^3/45$ and $L(x)_{x \gg 1} \approx 1 - 1/x$, we can rewrite the Eq. (52) for the case of $A\xi \ll 1$:

$$\left.\frac{\chi_T}{n}\right|_{A\xi \ll 1} = \frac{m_0^2}{k_B T}\frac{A^2}{3}\left(1 - \frac{1}{5}(A\xi)^2\right) + (1-A)\frac{2\mu+1}{2\mu-2}b^3. \qquad (53)$$

and for the case of $A\xi \gg 1$:

$$\left.\frac{\chi_T}{n}\right|_{A\xi \gg 1} = \frac{m_0^2}{k_B T}\frac{1}{\xi^2} + (1-A)\frac{2\mu+1}{2\mu-2}b^3, \qquad (54)$$

It is convenient to introduce the fraction of the magnetic phase in the system:

$$f_m = \frac{NV_p}{V} = nV_p, \qquad (55)$$

where $V$ is the total system volume, $N$ is the number of the particles in the system, $V_p = 4\pi b^3/3$ is the particle volume. Fig. 6. shows the normalized dependencies of isothermal differential magnetic susceptibility $\chi_T/f_m$ on $\xi$ for different values of $\mu$ and $A$. As the external magnetic field increases, the isothermal differential susceptibility decreases monotonously and tends to the following limiting value (see Eqs.(55), (54)):

$$\left.\frac{\chi_T}{f_m}\right|_{H_0 \gg 1} = \frac{3}{4\pi}(1-A)\frac{2\mu+1}{2\mu-2}, \qquad (56)$$

If $\mu \gg 1$, Eq.(56) can be written as:

$$\chi_T|_{H_0 \gg 1, \mu \gg 1} = (1-A)nb^3. \tag{57}$$

In this case, the value of parameter $A$ can be easily determined, if high-field $\chi_T$, the particle size $b$ and the particle concentration $n$ are known.

## 5. Discussion

The screening factor $A$ introduced in formula (20) turned out to be a very convenient parameter for describing the screening magnetic effects for CSHS particles. This coefficient is included in formulas (21)-(23) for determining the magnetic field strength in the core and shell of the CSHS particle, as well as in the external space. The total magnetic moment of the particle (25) is equal to the sum of two terms, one of which is constant (independent of the external field), and the second one is proportional to $H_0$ (induced). The constant term represents the screened magnetic moment of the core, reduced by $A$. The role of the induced term increases as the screening effect of core increases (as $A$ decreases). The energy of the CSHS particle in the external field also includes the factor $A$ (see Eq. (33)). The partition function $Z$ in the Langevin model (Eq.(44)) is changed for CSHS particles in such a way that its argument is multiplied by $A$.

The average magnetic moment for the CSHS particles in the Langevin model is equal to the sum of the Langevin term and the induced one. The Langevin term is modified due to the screening effect using the factor $A$. The induced term does not depend on the core magnetization (Eq.(45)).

Several approaches can be used to determine the screening coefficient $A$ based on experimental data. One approach can use the features of the isothermal field dependences of the magnetic moment (Fig. 4a). Another method is based on the peculiarities of the deviation of the magnetic moment temperature dependences from the Curie law (Fig. 5). In the third method, the parameter $A$ can be determined using the susceptibility value in high magnetic fields (Fig. 6).

The Langevin superparamagnetism model has drawbacks. For example, it does not take into account the anisotropy of the core material. However, it is shown in [48] that for uniformly distributed directions of the anisotropy axes in a single-domain nanoparticle system, the difference between average magnetic moments in the systems with and without anisotropy is not too large (less than 2%, Fig. 11 in [48]).

It should also be noted that in our model, the magnetic permeability is constant. This is obvious simplification for bulk soft ferromagnets, which magnetic susceptibility depends on the external magnetic field. However, in relation to nanoparticles it can be partially justified, since many types of magnetically soft nano-materials (for example, Fe-Ni, $\gamma-Fe_2O_3$, Ni, Si-Fe) demonstrate a linear dependence of the magnetization $M$ on $H$ for a larger range of external fields compared with similar bulk materials [55-60].

Numerous modern advanced methods for producing of nanoparticles (see, for example, [17-18]) could be applied to create CSHS nanoparticles. Our theoretical results can be useful to predict and control their magnetic properties. For example, weakening or strengthening of magnetic interparticle interactions and the interaction of the particle with external magnetic field can be critically important in the directed drug delivery methods. Enhancing the screening effect can prevent undesirable agglomerations of the particles induced by dipole – dipole magnetic interactions. Attenuating the screening effect can be useful for increasing drag force exerted on the particle by an external magnetic field in these methods. The ability to control the effects of screening is due to the dependence of the parameter $A$ on the magnetic permeability $\mu$, the value of which can be controlled using an external magnetic field.

## 6. Conclusion

The main purpose of our work was to study possible magnetic screening effects in nano-objects. Magnetostatic properties of CSHS particles were analytically outlined. As a result of solving the Laplace equation, analytical expressions were obtained for both the magnetic potential and the magnetic field strength in the entire space. In our work, we showed that the magnetically soft surface layer screens both the external space from the magnetic field of the core and the core from the external magnetic field. The screening effects can be useful for the applications, in which magnetic properties of nanoparticle should be controlled.

*This research did not receive any specific grant from funding agencies in the public, commercial, or not-for-profit sectors.*

**Acknowledgement**

**Declaration of Competing Interest**

The authors declare that they have no known competing financial interests or personal relationships that could have appeared to influence the work reported in this paper.

**Appendix A**

In this Appendix, we present the three independent sets of linear equations and their solutions for the constant coefficients $A_{10}^I, A_{11}^I, B_{11}^I, A_{10}^{II}, A_{11}^{II}, B_{11}^{II}, C_{10}^{II}, C_{11}^{II}, D_{11}^{II}, A_{10}^{III}, A_{11}^{III}, B_{11}^{III}$.

The set of equations in the variables $A_{10}^I, A_{10}^{II}, A_{10}^{III}, C_{10}^{II}$:

$$\begin{cases} A_{10}^I = \dfrac{b^3 - a^3}{a^3} C_{10}^{II} \\ \dfrac{m_0}{b^3} + \dfrac{b^3 - a^3}{b^3} A_{10}^{II} = -H_{0z} + \dfrac{1}{b^3} A_{10}^{III} \\ \dfrac{2m_0}{a^3}(1 - \mu) - A_{10}^I = -3\mu A_{10}^{II} + \mu \dfrac{2b^3 + a^3}{a^3} C_{10}^{II} \\ -2\mu \dfrac{m_0}{b^3} + \mu \dfrac{2a^3 + b^3}{b^3} A_{10}^{II} - 3\mu C_{10}^{II} = -H_{0z} - \dfrac{2}{b^3} A_{10}^{III} \end{cases} \quad (A.1)$$

The set of equations in the variables $A_{11}^I, A_{11}^{II}, A_{11}^{III}, C_{11}^{II}$:

$$\begin{cases} A_{11}^I = \dfrac{b^3 - a^3}{a^3} C_{11}^{II} \\ \dfrac{b^3 - a^3}{b^3} A_{11}^{II} = -H_{0x} + \dfrac{1}{b^3} A_{11}^{III} \\ -A_{11}^I = -3\mu A_{11}^{II} + \mu \dfrac{2b^3 + a^3}{a^3} C_{11}^{II} \\ \mu \dfrac{2a^3 + b^3}{b^3} A_{11}^{II} - 3\mu C_{11}^{II} = -H_{0x} - \dfrac{2}{b^3} A_{11}^{III} \end{cases} \quad (A.2)$$

The set of equations in the variables $B_{11}^I, B_{11}^{II}, B_{11}^{III}, D_{11}^{II}$:

$$\begin{cases} B_{11}^{I} = \dfrac{b^3 - a^3}{a^3} D_{11}^{II} \\ \dfrac{b^3 - a^3}{b^3} B_{11}^{II} = -H_{0y} + \dfrac{1}{b^3} B_{11}^{III} \\ -B_{11}^{I} = -3\mu B_{11}^{II} + \mu \dfrac{2b^3 + a^3}{a^3} D_{11}^{II} \\ \mu \dfrac{2a^3 + b^3}{b^3} B_{11}^{II} - 3\mu D_{11}^{II} = -H_{0y} - \dfrac{2}{b^3} B_{11}^{III} \end{cases} \quad (A.3)$$

The expressions for constant coefficients $A_{10}^{I}$, $A_{11}^{I}$, $B_{11}^{I}$, $A_{10}^{II}$, $A_{11}^{II}$, $B_{11}^{II}$, $C_{10}^{II}$, $C_{11}^{II}$, $D_{11}^{II}$, $A_{10}^{III}$, $A_{11}^{III}$, $B_{11}^{III}$ resulting from equations (A.1) – (A.3):

In region I ($r < a$):

$$A_{10}^{I} = -A H_{0z} - \frac{m_0}{a^3}(1 - A)\frac{\mu + 2}{\mu - 1}; \quad (A.4)$$

$$A_{11}^{I} = -A H_{0x}; \quad (A.5)$$

$$B_{11}^{I} = -A H_{0y}; \quad (A.6)$$

In region II ($a < r < b$):

$$A_{10}^{II} = -A\left(\frac{1}{3}\left(\frac{2b^3 + a^3}{b^3 - a^3} + \frac{1}{\mu}\right)H_{0z} + \frac{2(\mu - 1)^2}{9\mu b^3} m_0\right); \quad (A.7)$$

$$A_{11}^{II} = -\frac{A}{3}\left(\frac{2b^3 + a^3}{b^3 - a^3} + \frac{1}{\mu}\right)H_{0x}; \quad (A.8)$$

$$B_{11}^{II} = -\frac{A}{3}\left(\frac{2b^3 + a^3}{b^3 - a^3} + \frac{1}{\mu}\right)H_{0y}; \quad (A.9)$$

$$C_{10}^{II} = -A\left(\frac{a^3}{(b^3 - a^3)}H_{0z} + \frac{2(\mu - 1)(\mu + 2)}{9\mu b^3} m_0\right); \quad (A.10)$$

$$C_{11}^{II} = -A\frac{a^3}{b^3 - a^3}H_{0x}; \quad (A.11)$$

$$D_{11}^{II} = -A\frac{a^3}{b^3 - a^3}H_{0y}; \quad (A.12)$$

In region III ($r > b$):

$$A_{10}^{III} = A m_0 + (1 - A)\frac{2\mu + 1}{2\mu - 2}b^3 H_{0z}; \quad (A.13)$$

$$A_{11}^{III} = (1-A)\frac{2\mu+1}{2\mu-2}b^3 H_{0x}; \quad (A.14)$$

$$B_{11}^{III} = (1-A)\frac{2\mu+1}{2\mu-2}b^3 H_{0y}; \quad (A.15)$$

**Appendix B**

Let us consider the following special cases to make sure that the formulas (21) and (23) are correct.

- the field at the particle center (in the region I) in the absence of core magnetic moment $\vec{m}_0$:

$$\vec{H}_I\big|_{m_0=0} = A\vec{H}_0 \quad (B.1)$$

- the field on the outer surface of the particle (in the region III) in the absence of an external field $\vec{H}_0$:

$$\vec{H}_{III}\big|_{H_0=0} = A\left(\frac{3\vec{r}(\vec{m}_0\vec{r}) - \vec{m}_0 r^2}{r^5}\right) = A\vec{H}_{core} \quad (B.2)$$

Equation (B.1) coincide with Eq.(5.122) obtained in the classical textbook [58]. The role of parameter A is to create the screening effect for the external magnetic field (Eq.(B.1)) inside the shell and for the magnetic field of the core outside the particle (Eq.(B.2)).

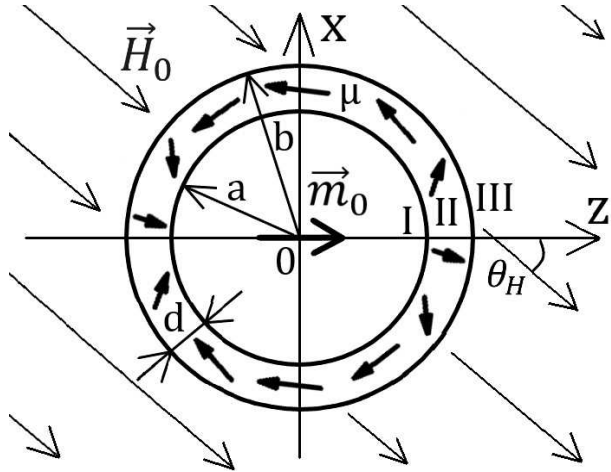

FIG. 1. The model of a core – shell hard – soft (CSHS) spherical nanoparticle in the external homogeneous magnetic field $\vec{H}_0$. (I): a single-domain hard ferromagnetic core of radius $a$; (II): a homogenous soft magnetic shell of external radius $b$ and thickness $d$; (III): outer nonmagnetic space. $\theta_H$ - the angle between the magnetic moment $\vec{m}_0$ of the core and $\vec{H}_0$.

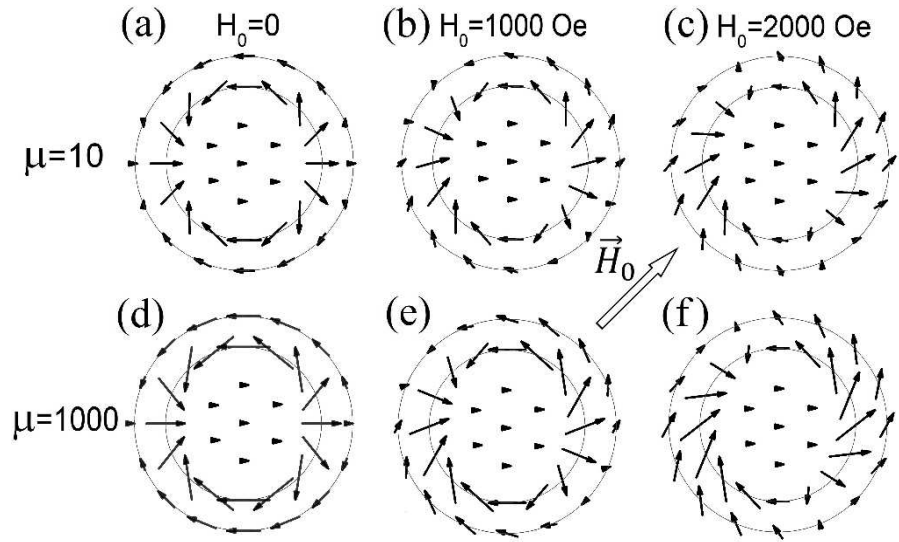

Fig. 2. The distribution of magnetization (indicated by arrows) at the inner and outer surfaces of the CSHS particle shell. The angle between $\vec{M_0}$ and $\vec{H_0}$ is equal to $\pi/4$. The length of the arrow is proportional to the magnetization magnitude. The value of $\mu$ is equal to 10 for (a)-(c) and 1000 for (d)-(f). The value of $H_0$ is equal to zero for (a),(d); 1 kOe for (b),(e); 2 kOe for (c),(f). Other model parameters: $a = 10\ nm;\ b/a = 1{,}4;\ M_0 = 1\ kG;\ T = 298\ K$.

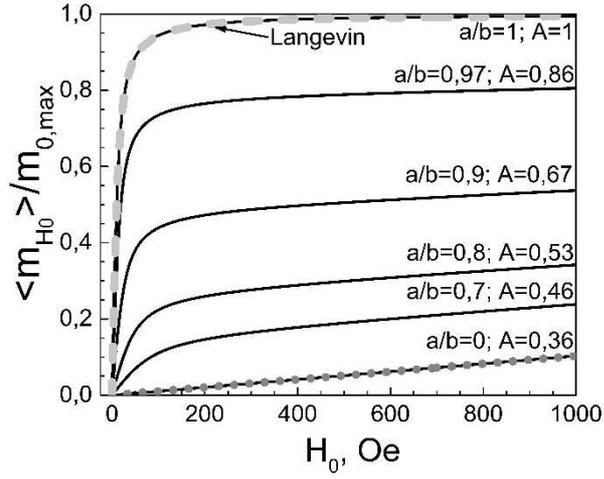

FIG. 3. The average magnetic moment along the direction of the external field $\langle m_{\vec{H}_0} \rangle$ as a function of the external field strength $H_0$ for different values of ratio $a/b$ (parameter $A$). The values of $\langle m_{\vec{H}_0} \rangle$ are normalized to $m_{0,max}$ (see Eq. (48)). Solid lines were calculated by using (45) taking $b = 10\ nm; \mu = 10; M_0 = 1{,}7\ G; T = 298$. The dashed line and dotted lines are drawn using Eq. (46) ($b = a$) and Eq. (47) ($a = 0$), respectively.

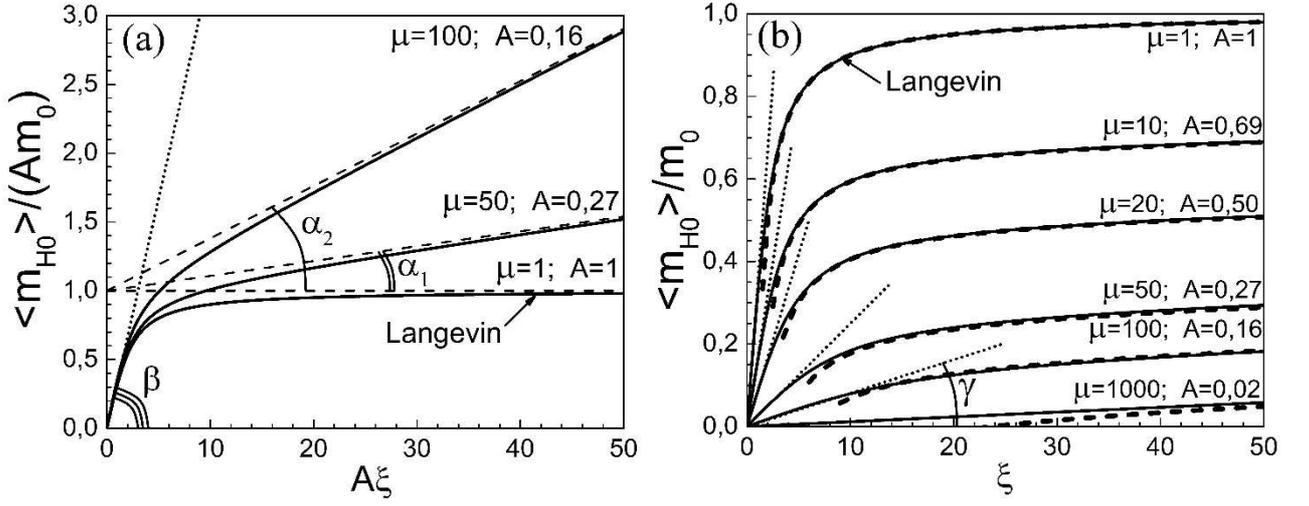

FIG. 4. The average magnetic moment $\langle m_{\vec{H}_0}\rangle$ as a function of (a) $\xi$ and (b) $A\xi$ for different values of shell permeability $\mu$ (parameter $A$). The values of $\langle m_{\vec{H}_0}\rangle$ are normalized to (a) $Am_0$ and (b) $m_0$. Solid lines were calculated by using (45) taking $a = 10\ nm$; $b/a = 1,1$; $M_0 = 1,7\ kG$; $T = 298\ K$. The dotted and dashed lines represent the tangents to asymptotes for the cases $A\xi \ll 1$ (Eq. (49)) and $A\xi \gg 1$ (Eq. (50)), respectively. The thick dashed lines in (b) represent the asymptotes for the cases $A\xi \gg 1$ (Eq. (50)). The angles $\alpha_1, \alpha_2, \beta, \gamma$ indicate the slope of the tangents to the asymptotes.

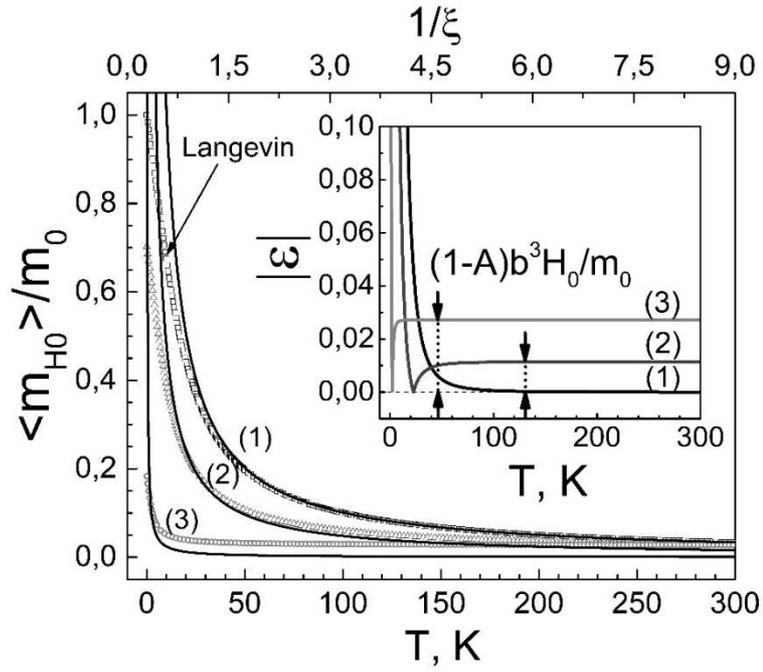

FIG. 5. The average magnetic moment $\langle m_{\vec{H}_0} \rangle$ as a function of $T$ ($1/\xi$) for different values of shell permeability $\mu$ (parameter $A$). The dotted lines were calculated by using Eq. (45). Solid lines were calculated by using Eq. (51). The inset shows the deviation modulus of the average magnetic moment from the Curie law $\varepsilon = (m_c - \langle m_{\vec{H}_0} \rangle)/m_0$ versus $T$. The values of $\langle m_{\vec{H}_0} \rangle$ are normalized to $m_0$. The lines (1) – (3) were calculated by using $H_0 = 10\ Oe$; $M_0 = 100\ G$; $a = 10\ nm$; $b/a = 1,1$; (1): $\mu = 1$ ($A = 1$); (2): $\mu = 10$ ($A = 0,69$); (3): $\mu = 100$ ($A = 0,16$). The formula shown in the inset is valid for $\mu > 10$.

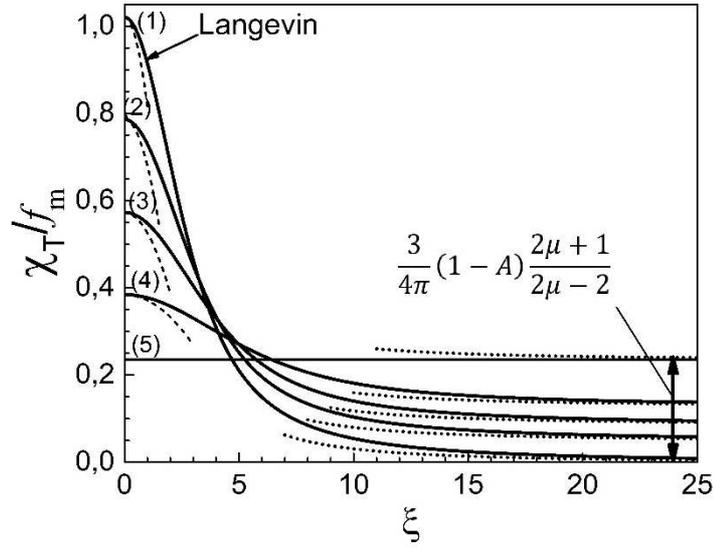

FIG. 6. The isothermal differential magnetic susceptibility $\chi_T$ as a function of the normalized external field $\xi$ for different values of $\mu$ and $A$. The values of $\chi_T$ are normalized to the magnetic component fraction $f_m$. Solid lines were calculated by using (52) taking $a = 10\ nm$; $b/a = 1,1$; $M_0 = 0,2\ kG$; $T = 298\ K$; (1): $\mu = 1\ (A = 1)$; (2): $\mu = 5\ (A = 0,85)$; (3): $\mu = 10\ (A = 0,69)$; (4): $\mu = 20\ (A = 0,5)$; (5): $\mu = 1000\ (A = 0,02)$. The dashed and dotted lines represent the asymptotes for the cases $A\xi \ll 1$ (Eq. (53)) and $A\xi \gg 1$ (Eq. (54)), respectively. The presented formula describes the deviation of $\chi_T/f_m$ from zero at high fields (see Eq. (54)).